\documentclass[preprint,3p]{elsarticle}
\usepackage{url}

\def\options{29}

\def\entry#1#2{\parbox[t]{5.5cm}{\it  #1:}\hspace*{0.35cm}\parbox[t]{11.0cm}{#2}\\[0.1cm]}

\begin{document}

\title{Top++: a program for the calculation of the top-pair cross-section at hadron colliders}

\author[Aachen]{Micha\l{}  Czakon}
\author[CERN]{Alexander Mitov}

\address[Aachen]{Institut f\"ur Theoretische Teilchenphysik und Kosmologie,
RWTH Aachen University, D-52056 Aachen, Germany}
\address[CERN]{Theory Division, CERN, CH-1211 Geneva 23, Switzerland}

\date{\today}

\cortext[thanks]{Preprint numbers: CERN-PH-TH/2011-303, TTK-11-58}

\begin{abstract}
We present the program {\tt Top++} for the numerical evaluation of the total inclusive cross-section for producing top quark pairs at hadron colliders. The program calculates the cross-section in a) fixed order approach with exact next-to-next-to leading order (NNLO) accuracy and b) by including soft-gluon resummation for the hadronic cross-section in Mellin space with full next-to-next-to-leading logarithmic (NNLL)  accuracy. The program offers the user significant flexibility through the large number (\options) of available options. {\tt Top++} is written in C++. It has a very simple to use interface that is intuitive and directly reflects the physics. The running of the program requires no programing experience from the user.
\end{abstract}

\maketitle

\section*{Program summary}

\entry{Name of the program}{Top++ (ver.~2.0).}

\entry{Program's homepage}{\tt  \url{http://www.alexandermitov.com/software}}

\entry{License, Warranty}{ GNU Public License. No warranty given or implied.}

\entry{Compiler}{ Developed and tested with GNU Compiler Collection's C++ compiler.}

\entry{Operating system}{  Linux; Mac OS X; can be adapted for other unix systems.}

\entry{Program language}{ C++.}

\entry{Memory required to execute}{ Typically less than 200 MB.}

\entry{External libraries}{ GNU Scientific Library (GSL); the Les Houches Accord pdf Interface (LHAPDF).}

\entry{Keywords}{  Top-quark, Resummation, QCD, Precision Physics, Hadron Colliers.}

\entry{Typical running time}{  Depending on the options. The program is optimized for speed.} 

\entry{Accuracy}{ Sub per-mill accuracy achievable in realistic time (program does not employ Monte Carlo methods).}

\section{Introduction: what is this program for?}

The program {\tt Top++} calculates the total inclusive top pair production cross-section in hadronic collisions. The program can be used in both pure fixed order perturbation theory through exact next-to-next-to leading order (NNLO) \cite{Czakon:2013goa,Czakon:2012pz,Czakon:2012zr,Baernreuther:2012ws} and by including soft-gluon resummation through next-to-next-to-leading logarithmic order (NNLL) \cite{Beneke:2009rj,Czakon:2009zw} matched through NNLO. The implementation of the soft gluon resummation is as in Ref.~\cite{Cacciari:2011hy}. {\tt Top++} is the first publicly available program that can perform soft gluon resummation in top-pair production. 
\footnote{After {\tt Top++} was released, the program {\it TOPIXS} \cite{Beneke:2012wb} appeared. It performs NNLL resummed calculations in $x$-space. The program {\it HATHOR} \cite{arXiv:1007.1327} performs calculations in fixed order perturbation theory.}

The program is written in C++ in a modern, modular and object-oriented way. It should be very easy to install on most Linux systems; please consult \ref{sec:install} for details. Moreover, once installed, the program is trivial to run. The program has been written with a user in mind that has no programming experience whatsoever. For that reason, the program has a very simple user interface that is the only point of contact between the program and the user. The user interface is described in Section \ref{sec:user-interface}. Users with average programming experience will find the program very easy to customize to their own needs. That may not be needed, however, since all options that have arisen in our own work on the subject have already been pre-programmed and are easy to access directly through the user interface.

In this manual we do not describe the physics in detail. A short description, needed to make the reading of the program's options self-consistent, can be found in Section \ref{sec:physics}. The relevant description of the physics and the options implemented in this program can be found in Refs.~\cite{Czakon:2013goa,Czakon:2012pz,Czakon:2012zr,Baernreuther:2012ws, Cacciari:2011hy}. The results in these papers represent the most advanced results in top physics to date.

And a word of caution regarding the numerical results derived with the help of the program. Settings, mainly related to the workings of the program, can affect the final numerical value (through numerical precision) but that should be at a precision level below what is physically relevant. As a rule of thumb we recommend producing numbers that are correct at the level of one per-mill (i.e. ${\cal O}(10^{-3})$). Precision better than that is a pure matter of taste on the side of the user (and, at a certain level, of the ability of the otherwise very capable integration routines).

\section{How to use the program}\label{sec:user-interface}

In the following we assume that the program has been installed and is running correctly; see ~\ref{sec:install} for more details. Next, we assume that the program is installed in a directory called ${\tt top++}$ and we are already there, i.e. the command line reads:
\footnote{Please note that the precise text on the left of the cursor depends on the particular terminal.}
\begin{verbatim}
~ top++$
\end{verbatim} 
Next one needs to open the file {\sl top++.cfg} in a text editor. Using, for example, {\sl pico} one types:
\begin{verbatim}
~ top++$ pico top++.cfg
\end{verbatim} 
Once the user has set all options at their desired values, the file needs to be saved. A number of examples are supplied with the program; see ~\ref{sec:examples}. Then the user needs to execute the program by typing:
\begin{verbatim}
~ top++$ ./top++
\end{verbatim} 

The program starts running, displays its step-by-step progress, timing pre-defined milestones. Once the run is completed, the final result is displayed and the program exits. The output can be found on the screen and in the file {\sl top++.res} located in the program's directory. If a new run is desired, one has to simply repeat all the steps described above. 

In the following we describe all \options\,options that are available to the user through the file {\sl top++.cfg}. The options are grouped into five subgroups. Please note that since all options have predefined default values, the user needs to only specify the values for options that differ from their default values. In particular, if all options are at their default values the file {\sl top++.cfg} can be empty. It might be convenient to keep certain options typed in the file, but when not in use the user can comment them out by putting the symbol {\sl /} at the beginning of a line.
\begin{enumerate}
\item {\bf General Setup} (type of collider, pdf set, pure fixed order calculation versus one with resummation).\\
\begin{enumerate}
\item {\tt Collider:}~ Takes two values: {\tt TEV} (default) or {\tt LHC} as labels for $\bar p p$ or $pp$ colliders.\\
\item {\tt  WithResummation:}~ Takes two values: {\tt YES} (default) and {\tt NO}. If set to {\tt YES} the program will compute the observable by including soft-gluon resummation with $N^nLO + N^kLL$ accuracy. The values of $n,k$ are set through the parameters in the subgroup {\bf Resummation}, below. In this case all settings in the subgroup {\bf Fixed Order} become irrelevant. If set to {\tt NO} then no soft gluon resummation is performed, i.e. the calculation is done at fixed order. In such a case the calculation is controlled through the options in the group {\bf Fixed Order}, and all options in the group {\bf Resummation} become irrelevant.\\
\item {\tt PDFuncertainty:}~ If set to {\tt YES} the program will compute and display the pdf uncertainty. When the pdf uncertainty is being calculated, central scales choice $\mu_F=\mu_R=m_t$ is set automatically, independently of the values for these scales set by the user (as described below). If {\tt  PDFuncertainty} is set to {\tt NO} (default) then the program computes and displays the scale variation for the choice of $\mu_F,\mu_R$ specified by the user (explained below) for a single pdf member (specified by the user; see option \ref{item:pdfmem} below). Detailed information about computing pdf uncertainties can be found  in section \ref{pdfuncertainty}.\\
\item {\tt RestrictedScaleVariation:}~ Takes either {\tt NO} or a number greater than or equal to $1.0$. This option allows the user to set any restriction (or no restriction at all if {\tt RestrictedScaleVariation NO}) on the allowed ratio for the renormalization and factorization scales. For more information see the description of the group of options {\bf Top quark mass and renormalization/factorization scales}. The default value is {\tt 2.0} which corresponds to the restricted scale variation of Refs.~\cite{Czakon:2013goa,Czakon:2012pz,Czakon:2012zr,Baernreuther:2012ws, Cacciari:2011hy}.\\ 
\item {\tt PDFset:}~ The pdf set. The program uses the LHAPDF library~\cite{Whalley:2005nh} and follows its nomenclature. The default set is {\tt MSTW2008nnlo68cl} \cite{Martin:2009iq}. Both ``.LHgrid" and ``.LHpdf" files can be used, see the option {\tt PdfFileType} below.\\
\item {\tt PDFmember:}\label{item:pdfmem}~ The specific member of the pdf set {\tt PDFset} that the user wishes to use for the calculation of the cross-section. We assume that the counting of pdf members starts from ``0". Default value is {\tt 0}.\\
\end{enumerate}	
\item {\bf Top quark mass and renormalization/factorization scales}. Please note that the ranges of {\tt muR} and {\tt muF} (defined in the following) need not be equal in length. The range of their ratio is unrestricted if {\tt RestrictedScaleVariation NO} is chosen. When the option {\tt RestrictedScaleVariation} takes a numerical value (with the constraint ${\tt RestrictedScaleVariation} \geq 1.0$), then the ratio of the two scales is restricted between: 
$${1\over {\tt RestrictedScaleVariation}} \le \mu_F/\mu_R \le {\tt RestrictedScaleVariation} \, . $$\\
\begin{enumerate}
\item {\tt Mtop:}~ The value of the on-shell top mass (in GeV). If the user would like to loop over a range of values for the top mass (see the description of the following two options) {\tt Mtop} represents the lower end of that range. Default value is {\tt 173.3}.
\item {\tt MtopLimit:}~ The upper limit of the range of values for the top mass that the user wants to loop over. If {\tt MtopLimit} $<$ {\tt Mtop} then the program automatically sets {\tt MtopLimit} $=$ {\tt Mtop} and issues a warning message. The default is the unattainable value {\tt MtopLimit} = $-1$. The value $-1$ is converted internally to {\tt MtopLimit} $=$ {\tt Mtop} too, but is special, because only for it the (annoying) warning message is suppressed.
\item {\tt MtopStep:}~ A positive number (need not be an integer) that specifies the step with which the value of the top mass is incremented in a loop. Its default value is {\tt 1}. 
\item {\tt muR:}~ A set of values for the renormalization scale (in units of the top mass {\tt Mtop}). An arbitrary number of values is allowed. By default {\tt muR} takes the set of values 0.5 ~1.0~ 2.0. A fine scan can be achieved with the help of the following set: 
$${\tt muR ~0.50 ~0.55 ~0.60 ~0.65 ~0.70 ~0.75 ~0.80 ~0.85 ~0.90 ~0.95 ~1.0 ~1.1 ~1.2 ~1.3 ~1.4 ~1.5 ~1.6 ~1.7 ~1.8 ~1.9 ~2.0} \, .$$
The larger set takes much longer to compute and typically returns the same result as the restricted set consisting of three elements. We have encountered, however, exceptional situations that in our experience lead to differences of up to 0.5\%. 
\item {\tt muF:}~ A set of values for the factorization scale (in units of the top mass {\tt Mtop}). It is set and used independently of the renormalization scale described above. Its default value is 0.5 ~1.0~ 2.0 and can be set analogously to {\tt muR}. 
\end{enumerate}
\item {\bf Resummation} (all options in this group are irrelevant when {\tt WithResummation NO} is chosen).\\
\begin{enumerate}
\item {\tt OrderFO:}~ Takes values {\tt LO, NLO} or {\tt NNLO} (default). It specifies the fixed order accuracy of the resummed result as well as the order through which the resummed exponent is matched; see section \ref{sec:physics} for more details.\\
\item {\tt OrderRES:}~ Takes the values {\tt LL, NLL} and {\tt NNLL} (default). It specifies the logarithmic accuracy of the resummation;  see section \ref{sec:physics} for more details.\\
\item {\tt A:}~ The value of the parameter $A$ introduced in Ref.~\cite{Bonciani:1998vc}. Typically used with the value {\tt A 0} (default) or {\tt A 2}.\\
\item {\tt TwoLoopCoulombs:}~ Takes the values {\tt YES} (default) and {\tt NO}. It includes/excludes the two loop Coulombic terms in the function $\sigma^{\rm (Coul)}$, see Eq.~(\ref{eq:res}) below.  
\item {\tt H2qq, H2gg1, H2gg8:}\label{item:H2}~ The three constants in the two-loop hard function defined in a normalization of $\alpha_S/\pi$. Implemented is the two-loop matching derived in Refs.~\cite{Czakon:2013goa,Baernreuther:2012ws} (see also the description of the options  below): 
\begin{eqnarray}
&&{\tt H2qq} = 84.81\,\,\, ,\,\, {\tt H2gg1} =53.17\,\,\, ,\,\, {\tt H2gg8} =96.34\, .
\label{H2}
\end{eqnarray}
Please note that in addition to {\tt H2qq} only the color averaged combination of the $gg$-initial state matching coefficients {\tt H2gg1, H2gg8 } is presently known \cite{Czakon:2013goa}.\\
\end{enumerate}
\item {\bf Fixed Order}\label{FO} (all options in this group are irrelevant when {\tt WithResummation YES} is chosen).\\
\begin{enumerate}
\item {\tt LO:}~ takes values {\tt YES} (default) or {\tt NO}.\\
\item {\tt NLO:}~ takes values {\tt YES} (default) or {\tt NO}.\\
\item {\tt NNLO:}~ takes values {\tt YES} (default) or {\tt NO}. This option implements the exact NNLO result for $t\bar t$ production at hadron colliders \cite{Czakon:2013goa,Czakon:2012pz,Czakon:2012zr,Baernreuther:2012ws}.\\
\end{enumerate}
The options in this group allow access to each individual order in perturbation theory. For example, the option {\tt NLO} controls only the term $\sim \alpha_s^3$, i.e. for a calculation with NLO accuracy one has to set {\tt LO YES, NLO YES} and {\tt NNLO NO}. 
\item {\bf Setup parameters} (parameters related to the working of the program and other, less frequently modifiable parameters).\\
\begin{enumerate}
\item {\tt ECMLHC:}~ The c.m. energy of the $pp$ collider (in GeV). The default value is {\tt 8000}.\\
\item {\tt ECMTEV:}~ The c.m. energy of the $\bar p p$ collider (in GeV). The default value is {\tt 1960}.\\
\item {\tt Precision:}~ Defines the required relative precision of the integration routines: $${\rm Relative} ~ {\rm precision} = 10^{- {\tt Precision}}\, .$$ The default  value {\tt precision 2} tends to produce  fast and accurate (at the per-mill level) results.\\
\item {\tt NPdfGrid:}~ Defines the size of the grid on which the pdf fluxes are being discretized. The default value {\tt NPdfGrid 100} tends to produce fast and sufficiently accurate results.\\
\item {\tt  ETA:}~ Parameter introduced in Ref.~\cite{Catani:1996yz} that controls the subtraction flux implemented in the resummed calculation. The default value {\tt  ETA 1e-5} is optimal.
\footnote{We use the usual notation: $1e$-$p\equiv 10^{-p}$.}
The user normally will not need to be concerned with this parameter.\\
\item {\tt CMP:}~ Within the Minimal Prescription of Ref.~\cite{Catani:1996yz}, this option corresponds to the point where the contour for the inverse Mellin transform crosses the real line. The default value {\tt CMP 2.7} is optimal. The user normally need not be concerned with this parameter.\\
\item {\tt PdfFileType:}~ Option for switching between ``.LHgrid" and ``.LHpdf" files in the LHAPDF interface \cite{Whalley:2005nh}. The default value is {\tt LHgrid}. The alternative value is {\tt LHpdf}.\\
\item {\tt PartonChannel:}~ This option allows the user to compute the contribution of a single partonic channel to the total cross-section. It can be used with both fixed order and resummed calculations. To select a particular channel one has to select one of the following six values: {\tt qqbar, gg, qg, qq, qqprime, qqbarprime}. The default value is {\tt ALL} which represents the only phenomenologically relevant case when all partonic channels are included (i.e. {\tt ALL} is equivalent to the sum of the six partonic channels). Any other value, different from the six partonic channels is equivalent to {\tt ALL}. Needless to say, results derived from a single partonic channel have to be interpreted with care.
\end{enumerate}
\end{enumerate}

\section{Computation of pdf uncertainties}\label{pdfuncertainty}

Starting from ver.2.0, {\tt Top++} implements a new approach for the computation of pdf uncertainties. This approach is meant to guarantee  compatibility of the program with respect to future modifications in the {\it LHAPDF} interface \cite{Whalley:2005nh} or in existing families of pdf sets.  The basic logic is as follows: 
\begin{enumerate}
\item Four pdf computation prescriptions (called {\it Asymmetric}, {\it NNPDF}, {\it Symmetric}  and {\it HERA\_VAR}) are pre-built into the program. If needed, more prescriptions can be added by the user (see section \ref{addpdf} for details).
\item In order for the program to automatically compute pdf uncertainty for a specific pdf set, the full name of this particular pdf set has to be added to a library which associates it with the appropriate prescription for computing pdf uncertainty. Note that no default naming conventions are allowed, i.e. each pdf set has to be specified by its full name (without the extension). The program comes with a database of pairs $(pdf~set;~prescription)$ contained in the file {\tt pdf.cfg}. The default version of the file {\tt pdf.cfg} contains a number of five-flavor pdf sets, and more sets can be freely added by the user (in no particular order). Note that if this file is modified, the program need not be recompiled.
\end{enumerate}

If no known pdf prescription exists (in the file  {\tt pdf.cfg}) for the pdf set requested by the user (in the file {\tt top++.cfg}) then the program will not compute pdf uncertainty at all and a warning message will be displayed. Still, the results for all individual pdf members are computed and displayed, i.e. if no prescription for computing pdf uncertainty with the current pdf set is known, the displayed individual results can still be used by the user for further manual processing. 

The four available methods for computing pdf uncertainties are: 
\begin{enumerate}
\item The asymmetric prescription {\it Asymmetric} of Ref.~\cite{hep-ph/0110378} (see also Ref.~\cite{Cacciari:2008zb}). It is typically used for the MSTW \cite{Martin:2009iq} (and earlier sets) as well as the CTEQ family of sets \cite{Lai:2010vv}.
\item The {\it NNPDF} prescription \cite{Ball:2011uy} for the NNPDF family of sets.
\item The symmetric prescription {\it Symmetric} (as defined in \cite{Alekhin:2012ig}). It is typically used for the ABM11 \cite{Alekhin:2012ig}, ABKM09 \cite{Alekhin:2009ni} and earlier Alekhin \cite{Alekhin:2002fv} family of pdf sets.
\item The prescription {\it HERA\_VAR} for computing pdf variation of the {\it \_VAR}-type of sets from the HERA family of pdf sets (see \cite{Whalley:2005nh}). Note that the complete pdf variation is a combination of the corresponding {\it \_VAR} and {\it \_EIG} sets; the pdf variation of the latter set can be computed with the {\it Asymmetric} prescription.
\end{enumerate}

\subsection{Adding new prescriptions for computing pdf uncertainties}\label{addpdf}

New prescriptions might be needed for certain pdf sets, or might be developed in the future. In such cases the program in its current form will not compute the pdf uncertainty but will display the results for all individual pdf members. It will be up to the user then to derive the pdf uncertainty from these results.

To avoid the repetition of this time consuming work, the user might wish to add his/her own prescription inside {\tt Top++} and automate the task. The addition of a new prescription for pdf uncertainty requires a rather straightforward three-step modification of the code:
\footnote{The authors will be happy to assist users with this task and possibly incorporate new pdf prescriptions that might be of interest to a broader audience into future official versions of the program.}
\begin{enumerate}
\item \label{function} Add a function that defines the prescription itself. It is placed at the end of the file {\tt Utilities.cpp}; the file {\tt Utilities.h} has to be modified accordingly. As a guidance, one could use the four existing functions for computing pdf uncertainties.
\item\label{prescription}The name of the new pdf prescription must be accounted for inside the file {\tt top++.cpp} in the two lines following the comment {\it Setup for the computation of pdf uncertainties}. 
\item In the file {\tt top++.cpp}, pair the new pdf uncertainty prescription (defined in step \ref{prescription}) with the new function that computes it (defined in step \ref{function}), following the comment {\it Compute scale/pdf variation for a fixed mtop}. 
\end{enumerate}

\section{Once the program is running: some fine tuning}

\subsection{Numerical precision and speed}\label{sec:precision}

As every program for numerical calculations, {\tt Top++} has its limits, too. In the following we discuss this, as well as ways to improve the accuracy and shorten the length of the runs. 

There are two places where speed (and therefore accuracy) can be
controlled. The first one is through the option {\tt Precision} of the
integration routines. In our own experience the value {\tt Precision
  2} is more than adequate to calculate the $t\bar t$ cross-section to
per-mill accuracy. Increasing the value of {\tt Precision} slows the
calculation down. The program is set in such a way, that if the
integration routines cannot reach the accuracy in a point, a warning
message is displayed specifying the relative error returned by the
integration routines. The presence of these messages, as such, is
harmless. They can be ignored if the displayed relative precision is
high enough. Rarely, the output might contain the symbol NaN, which
stands for Not-a-Number. NaNs are returned due to invalid numerical
operations such as division by zero, or operations, which produce
numbers out of the range of double precision. This is a sign of
numerical instability. We have taken the pragmatic approach of not
handling such exceptions in any specific way (which is very difficult
in practice). In case of a NaN result, the user should rerun the
calculation with increased requested accuracy.

We recommend that at least once the user does the calculation with a larger value for {\tt Precision} and verifies that the change in the result is beyond the required accuracy.

A second (and independent) source of numerical uncertainty is the size {\tt NPdfGrid} of the grid over which the partonic fluxes are approximated. We have implemented a second order finite difference scheme. The relative precision scales as $\sim1/{\tt NPdfGrid}^2$. In practice we have noticed that a value {\tt NPdfGrid 100} is more than adequate in terms of accuracy and produces very fast calculations. We recommend that at least once the user does the calculation with a larger value, say {\tt NPdfGrid 500}, and verifies that the change in the result is beyond the required accuracy.

Finally, the user should keep in mind that the overall numerical accuracy is a combination of the settings for {\tt Precision} and {\tt NPdfGrid}, i.e. increasing only one of them to an extreme may not have a net positive effect on the overall uncertainty but might lead to a significant slowdown.

\subsection{Outputting the results}

The program outputs on the screen all results as well as information about the timing of each step. The specifics depend on the requested options. At the end of the calculation, a summary of the final result is displayed, including the central value (i.e. the value corresponding to central scale choice, if requested) and the scale or pdf variation's absolute and relative values (when prescription for pdf uncertainty is known or when central value can be computed). The program also outputs the final result in a file {\tt top++.res} which is ready for plotting with the program {\it Gnuplot}. This is particularly useful for the case when the user requests a loop over a range of values of the top quark mass. In such case the result for each value is conveniently recorded in the file.
Please note that the file {\tt top++.res} is overwritten after each run.

\subsection{Modifying the default values of the parameters}

The default values for all \options\,parameters available to the user are set inside the function {\tt main()} located in the file {\tt top++.cpp}. All default settings can be modified by the user, although this is not recommended and should not be necessary.

\subsection{Parallelizing calculations} 

Practice shows that oftentimes one has to perform a large number of calculations with different parameters. Given that the run times vary significantly, it is natural {\it and very beneficial} to parallelize these calculations. To achieve such parallelization the user basically needs to have in a common directory only the executable {\tt top++} (common for all calculations), together with the two configurations files {\tt pdf.cfg} (also common for all calculations) and {\tt top++.cfg} (specific to each calculation).

\subsection{Additional}

The strong coupling constant is calculated at a scale $\mu_R$ through the LHAPDF interface \cite{Whalley:2005nh}. The calculation of the cross-section is performed in a scheme with $N_F=5$ active flavors. For consistency only pdf sets with $N_F=5$ active flavors should be used for scales above $m_{top}$.

\section{Contact with physics}\label{sec:physics}

The program can be used to compute the $t\bar t$ total inclusive cross-section $\sigma_{\rm tot}$ either in a pure fixed order perturbation theory through exact NNLO \cite{Czakon:2013goa,Czakon:2012pz,Czakon:2012zr,Baernreuther:2012ws} or including soft gluon resummation performed in Mellin space through NNLL \cite{Beneke:2009rj,Czakon:2009zw} and matched through NNLO:
\begin{equation} 
\sigma_{\rm tot}^{(n,k)} = \sigma_{\rm F.O.}^{(n)}  +  \left[\sigma_{\rm res}^{(n,k)} - \sigma_{\rm res}^{(n,k)}\vert_{\alpha_S^n} \right]\, ,
\label{eq:sigma-tot}
\end{equation}
where the labels $n$ ($k$) implicitly denote the fixed (logarithmic) order accuracy of the result.

Pure fixed order calculation can be achieved by setting the option {\tt WithResummation NO}. In this case the terms in the square bracket in Eq.~(\ref{eq:sigma-tot}) are absent and $n$ is controlled through the appropriate combination of the options {\tt LO, NLO} and {\tt NNLO} described in section \ref{sec:user-interface}.\ref{FO}. We remind the reader that these three options control separately the corrections at orders {\cal O}$(\alpha_S^n),~n=2,3,4$ in the fixed order result. This way the user has individual access to each one of the three known orders in the perturbative expansion of $\sigma_{\rm F.O.}$. For example, for calculations with NNLO accuracy, one has to set all three options {\tt LO, NLO, NNLO} to {\tt  YES}.

To perform soft-gluon resummation one has to include the terms in the square bracket in Eq.~(\ref{eq:sigma-tot}) by setting the option {\tt WithResummation YES}. The power $n$ has the same meaning as in the case of pure fixed order calculation, but is now controlled through the option {\tt OrderFO}. This option takes the value {\tt LO} (or {\tt NLO}, or {\tt NNLO}) which, unlike the case of pure fixed order calculations described above, includes all terms through order {\cal O}$(\alpha_S^n),~n=2$ (or $3$, or $4$). 

In Mellin $N$-space, the resummed partonic cross-section reads (see Ref.~\cite{Cacciari:2011hy} for complete description)
\begin{equation}                    
\label{eq:res} 
\sigma^{(n,k)}_{{\rm res~part};N,{\bf I}} = \sum_{{\bf I = 1,8}}
\sigma^{{\rm (Coul)},(n)}_{N,{\bf I}} \times \; 
\sigma^{{\rm (Hard)},(n)}_{N,{\bf I}} \times \; 
\Delta^{(k)}_{N,{\bf I}} \, .
\end{equation}

The function $\sigma^{{\rm (Coul)},(n)}$ in Eq.~(\ref{eq:res}) contains the Coulombic effects and has a known perturbative expansion through NNLO. The depth of the perturbative expansion of this function is set by the option {\tt OrderFO}. The user can also turn on or off the NNLO correction to this function through the option {\tt TwoLoopCoulombs} (normally it should not be modified). The expression for the function $\sigma^{\rm (Coul)}$ can be found in Ref.~\cite{Cacciari:2011hy}.

The function $\sigma^{{\rm (Hard)},(n)}$ in Eq.~(\ref{eq:res}) is an $N$-independent function. The depth of its perturbative expansion, in an expansion in $\alpha_S/\pi$, is set by the option {\tt OrderFO}. The one-loop corrections are known exactly \cite{Czakon:2008cx}. The two-loop corrections {\tt H2gg1, H2gg8, H2qq} are accessible through the options {\tt H2gg1, H2gg8, H2qq}: the $q\bar q$ reaction term {\tt H2qq} is known with sufficiently high accuracy \cite{Baernreuther:2012ws}, while only the color averaged combination of {\tt H2gg1, H2gg8} is currently known \cite{Czakon:2013goa}. The default values for {\tt H2gg1, H2gg8} from Ref.~\cite{Czakon:2013goa} are given in Eq.~(\ref{H2}). 

The function $\Delta^{(k)}$ in Eq.~(\ref{eq:res}), the so-called Sudakov exponent, contains the towers of LL, NLL and NNLL soft $\ln(N)$ logs. The user can request LL, NLL or NNLL logarithmic accuracy by setting the option {\tt OrderRES} to {\tt LL, NLL}, or {\tt NNLL} respectively. For example, for resummation with full NNLL accuracy one sets {\tt OrderRES NNLL}. 

As follows from the description of the functions $\sigma^{\rm (Coul)}$ and $\sigma^{\rm (Hard)}$ above, for a given fixed order accuracy {\tt OrderFO} the resummed cross-section $\sigma_{\rm res}^{(n,k)}$ is automatically matched through the same order {\tt OrderFO} and,  to avoid double counting, its perturbative expansion $\sigma_{\rm res}^{(n,k)}\vert_{\alpha_S^n}$  is subtracted also through order {\tt OrderFO}. 
\footnote{We note that starting from ver. 2.0, the matching of the resummed cross-section $\sigma_{\rm res}$ is internally controlled through the option {\tt OrderFO} and not through the option {\tt OrderRES} as was the case in the earlier versions of {\tt Top++}. The new setting corresponds to the standard matching conventions and is the natural choice when working with the exact NNLO result (unlike working with approximate NNLO as was the case in earlier versions). Of course, for typical applications like LO+LL, NLO+NLL and NNLO+NNLL, the two implementations are equivalent.}

No options related to approximate NNLO calculations are available in the program, since they are now obsolete and superseded by the exact NNLO results. If the user would like to use them anyway, then he/she will have to use the appropriate earlier version of {\tt Top++}.

\section{Summary}

We present the C++ program {\tt Top++} for the calculation of the total inclusive top-pair cross-section at hadron colliders. This is the first publicly available program capable of performing soft-gluon resummation for this collider observable. The program incorporates all currently available theoretical results: fixed order calculations through NNLO \cite{Czakon:2013goa,Czakon:2012pz,Czakon:2012zr,Baernreuther:2012ws} and soft gluon resummation through NNLL \cite{Beneke:2009rj,Czakon:2009zw}. The user has access to \options\,options which results in a great deal of flexibility and control over the calculation. In this manual we have given only a very short introduction to the physics behind our program. For further details the user should consult Ref.~\cite{Cacciari:2011hy} as well as Refs.~\cite{Czakon:2013goa,Czakon:2012pz,Czakon:2012zr,Baernreuther:2012ws}. 

The program is organized in a modular, object oriented way.  It is optimized for speed given the user's requirements for accuracy. Our experience shows that, depending on the chosen options, the run times can vary significantly. In practice the speed can be an issue only for resummed calculations, due to the integration of rapidly oscillating functions in the complex plane. For fixed order calculations the run times are very short.

\noindent
\section*{Acknowledgments}

We thank Matteo Cacciari, Michelangelo Mangano and Paolo Nason for numerous cross-checks and a very fruitful collaboration \cite{Cacciari:2011hy} that led to the creation of this software. We also thank Juan Rojo, Graeme Watt and Malgorzata Worek for many useful suggestions. The work of M.C. was supported by the Heisenberg and by the Gottfried Wilhelm Leibniz programmes 
of the Deutsche Forschungsgemeinschaft, and by the DFG Sonderforschungsbereich/Transregio 9 
``Computergest\"utzte Theoretische Teilchenphysik''.

\appendix

\section{Installation}\label{sec:install}

The program has been written in standard {\tt C++} and has been tested to correctly compile under the GNU compiler {\tt g++} from version 4 upwards. It requires two external libraries:

\begin{itemize}

\item {\bf GNU Scientific library}, which can be downloaded free of
  charge from

  {\tt http://www.gnu.org/s/gsl/}

  and is used for special functions and integration

\item {\bf the Les Houches Accord pdf Interface}, which can be
  downloaded free of charge from

  {\tt http://projects.hepforge.org/lhapdf/}

  and is used for the parton distribution functions

\end{itemize}

To setup the program for installation, it is necessary to set three
variables in the {\tt Makefile} contained in the {\tt Top++}
installation directory.

\begin{itemize}
\item {\bf \tt CXX} - c++ compiler. Specify the full path if necessary

\item {\bf \tt GSLDIR} - prefix directory for the gsl library, it is assumed
         that the library is in GSLDIR/lib and the include
         files in GSLDIR/include

\item {\bf \tt LHADIR} - prefix directory for the Les Houches pdf library, it is
         assumed that the library is in LHADIR/lib and the
         include files are in LHADIR/include/LHAPDF
\end{itemize}
The paths to the directories LHADIR and GSLDIR are determined automatically. The paths can also be set explicitly, the most standard  case being {\tt GSLDIR = /usr/local} and {\tt LHADIR = /usr/local}.

In the next step, it is sufficient to compile the code using
\begin{verbatim}
~ top++$ make
\end{verbatim} 
The program should compile without any error massages or warnings, and
is ready to use under the name {\tt top++}.

Alternatively to modifying the Makefile it is possible to compile
directly with
\begin{verbatim}
~ top++$ make CXX="user CXX value" GSLDIR="user GSLDIR value" LHADIR="user LHADIR value"
\end{verbatim} 
The quotation marks above are not necessary, unless the paths contain
spaces or other special values (as usual under unix).

\section{Examples (updated for ver.2.0 )}\label{sec:examples}

A number of examples can be found in the directory {\tt examples}. The user can copy/paste their content into the file {\tt top++.cfg} located in the program's main directory.
\begin{itemize}
\item The file {\tt top++complete-set.cfg} contains all options offered by the program. Most of these options are not needed in typical phenomenological studies and therefore will rarely be used.
\item The file {\tt top++pheno.cfg} contains all parameters needed in typical phenomenological studies.
\item The file {\tt top++best-precision.cfg} is a subset of {\tt top++pheno.cfg} and contains a set of options needed for typical applications, i.e. calculations with the ``best" available NNLO+NNLL precision.
\end{itemize}

In the following, we give few examples based on the file {\tt top++pheno.cfg}.
\vskip 2mm

{\bf Example 1:} If one executes the program with its default settings provided in the file {\tt top++.cfg} (which is equivalent to the file {\tt top++pheno.cfg}) one should get the best prediction of Ref.~\cite{Czakon:2013goa} for the LHC at 8 TeV (this result can also be obtained with an empty file {\tt top++.cfg} or, alternatively, one with all its options commented out): 
$${\tt sigma\_tot = 245.794 + 6.23992 (2.53868\%) - 8.41868 (3.4251\%) ~ [pb].}$$

{\bf Example 2:} With the file {\tt top++best-TEV.cfg} one can compute the scale variation of the cross-section corresponding to the best prediction of Ref.~\cite{Czakon:2013goa} for the Tevatron:
$${\tt  sigma\_tot = 7.1642 + 0.109671 (1.53082\%) - 0.199974 (2.79129\%) ~ [pb].}$$

{\bf Example 3:} With the file {\tt top++pdfvar-best-LHC7.cfg} one can compute the pdf variation of the cross-section corresponding to the best prediction of Ref.~\cite{Czakon:2013goa} for the LHC at 7 TeV:
$${\tt sigma\_tot =  172.025 + 4.7067 (2.73606\%) - 4.79784 (2.78904\%) ~[pb].}$$

{\bf Example 4:} To derive the scale variation variation for the NNLO fixed order prediction for the LHC at 14 TeV one can use the file {\tt top++NNLO-LHC14.cfg}:
$${\tt sigma\_tot = 932.959 + 31.7776 (3.4061\%) - 50.9738 (5.46367\%) ~[pb].}$$

Next we give few examples of less common calculations.
\vskip 2mm

{\bf Example 5:} To derive the contributions from the term $\sim\alpha_s^3$ (i.e. not including the contributions at $\sim\alpha_s^2$ and $\sim\alpha_s^4$ and no soft-gluon resummation) for central scale values $\mu_F=\mu_R=m_{\rm top}$ at the LHC at 7 TeV and with NNLO pdf, we use the file {\tt top++as3-LHC7.cfg}:
$${\tt sigma\_tot = 47.6818 ~[pb].}$$

{\bf Example 6:} To compute the cross-section with NNLO+LL precision at the LHC at 8 TeV, we use the file {\tt top++NNLO-LL-LHC8.cfg}:
$${\tt sigma\_tot = 244.283 + 8.18593 (3.351\%) - 14.4041 (5.89649\%) ~[pb].}$$

{\bf Example 7:} To compute the cross-section at NLO, with NLO pdf, for three different values of the top mass (chosen here to be close to its current best value and uncertainty) at the Tevatron, we utilize the looping option over the top mass (see the file {\tt top++mass-loop-TEV.cfg}):
\begin{eqnarray}
&&{\tt m_{\rm top}=172.3:~~sigma\_tot = 6.89431 + 0.367937 (5.33682\%) - 0.776566 (11.2639\%) ~[pb]\, ,}\nonumber\\
&&{\tt m_{\rm top}=173.3:~~sigma\_tot = 6.68166 + 0.356095 (5.32945\%) - 0.752499 (11.2622\%) ~[pb]\, , }\nonumber\\
&&{\tt m_{\rm top}=174.3:~~sigma\_tot = 6.47632 + 0.344724 (5.32283\%) - 0.72929 (11.2609\%) ~[pb]\, .}\nonumber
\end{eqnarray}

\section{What is new in ver.2.0: changes from ver.1.4}

\begin{itemize}
\item Added the exact NNLO result for the reaction $gg\to t\bar t+X$ \cite{Czakon:2013goa}. With this addition the program now contains the full set of NNLO QCD corrections to top pair production at hadron colliders.
\item Improved formatting of the output file {\tt top++.res}.
\item New approach to computing pdf uncertainties (see section \ref{pdfuncertainty} for details). 
\item Modified the evaluation of pdf uncertainties to accommodate pdf sets with different $\alpha_S$ for each pdf member.
\item Removed the options {\tt NNLOonORoff}, {\tt RESonORoff} and {\tt Cbargg} related to approximate NNLO \cite{Beneke:2009ye} approaches.
\item Replaced the NDE \cite{Nason:1987xz} parameterization of the NLO partonic cross-section with a fit \cite{arXiv:1007.1327} to the exact NLO result \cite{Czakon:2008ii}. 
\item New default value for the option {\tt A} (was $2$, now is $0$).
\item New default values for the options  {\tt H2gg1} and {\tt H2gg8} (both were $0$; for current values see Eq.~(\ref{H2})).

\item Matching of the resummed result is set by the option {\tt OrderFO} (was set by the option {\tt OrderRES}); see also section \ref{sec:physics}. The two are equivalent in the standard case when equal fixed order and logarithmic precision is required, i.e. LO+LL, NLO+NLL and NNLO+NNLL.
\item Added the prescription {\it HERA\_VAR} for computing pdf variation of the {\it \_VAR}-type of sets from the HERA family of pdf sets.
\item Added new option {\tt PartonChannel} which allows the user to select the contribution from a single partonic channel.
\item Rewritten manual to reflect the above modifications.
\end{itemize}

\section{Program's structure: a brief overview}\label{program}

The function {\tt main()} is located in the file {\tt top++.cpp}. The program consists of four classes that are initialized in the function {\tt main()}. The classes, listed in the order they are initialized, and their functionality are briefly described in the following.
\begin{enumerate}
\item {\tt Class PartonicFlux.} This class precomputes the partonic fluxes on a grid of {\tt NPdfGrid} points. One object of the class is created for each value of the factorization scale $\mu_F$. See also Section~\ref{sec:precision}.
\item {\tt Class FixedOrder.} This class represents the partonic fixed order cross-section. One object of this class is created for each combination of $(\mu_F,\mu_R)$. 
\item {\tt Class SubtrFlux.} This class implements a fake partonic flux that mirrors the actual partonic flux within a distance  ${\cal O}({\tt  ETA})$ from the partonic threshold. Our implementation follows Ref.~\cite{Catani:1996yz} where this flux was introduced and the need for it explained. Objects of class {\tt SubtrFlux} are created for each object of class {\tt PartonicFlux.}
\item {\tt Class Resummation.}  One  object of this class is created for each pair $(\mu_F,\mu_R)$ if a resummed calculation is requested by the user. Each object is constructed analytically in Mellin $N$-space and then inverted numerically back to $x$-space with the help of the Minimal Prescription of Ref.~\cite{Catani:1996yz}.
\end{enumerate}
A small number of functions can be found outside the above classes. The file {\tt lgamma} contains the logarithm of the Euler gamma function $\ln\Gamma(z)$ for complex argument, and {\tt psin} contains the polygamma function $\Psi(z,k),~ k=0,1$ for complex $z$. The remaining functions are located in the file {\tt Utilities}. These are the prescriptions for calculation of pdf uncertainties (we have implemented four prescriptions; see the description of option {\tt PDFuncertainty} in Section~\ref{sec:user-interface}) and the actual computation of the final result as a convolution of perturbative functions and partonic fluxes.

\end{document}